\documentclass[twoside,twocolumn,9pt]{article}
\usepackage{extsizes}
\usepackage[super,sort&compress,comma]{natbib} 
\usepackage[version=3]{mhchem}
\usepackage[left=1.5cm, right=1.5cm, top=1.785cm, bottom=2.0cm]{geometry}
\usepackage{balance}
\usepackage{times,mathptmx}
\usepackage{sectsty}
\usepackage{graphicx} 
\usepackage{lastpage}
\usepackage[format=plain,justification=raggedright,singlelinecheck=false,font={stretch=1.125,small,sf},labelfont=bf,labelsep=space]{caption}
\usepackage{float}
\usepackage{fancyhdr}
\usepackage{fnpos}
\usepackage[english]{babel}
\usepackage{array}
\usepackage{droidsans}
\usepackage{charter}
\usepackage[T1]{fontenc}
\usepackage[usenames,dvipsnames]{xcolor}
\usepackage{setspace}
\usepackage[compact]{titlesec}

\usepackage{epstopdf}

\usepackage{extarrows}

\definecolor{cream}{RGB}{222,217,201}

\begin{document}

\pagestyle{fancy}
\thispagestyle{plain}
\fancypagestyle{plain}{

\renewcommand{\headrulewidth}{0pt}
}

\makeFNbottom
\makeatletter
\renewcommand\LARGE{\@setfontsize\LARGE{15pt}{17}}
\renewcommand\Large{\@setfontsize\Large{12pt}{14}}
\renewcommand\large{\@setfontsize\large{10pt}{12}}
\renewcommand\footnotesize{\@setfontsize\footnotesize{7pt}{10}}
\makeatother

\renewcommand{\thefootnote}{\fnsymbol{footnote}}
\renewcommand\footnoterule{\vspace*{1pt}%
\color{cream}\hrule width 3.5in height 0.4pt \color{black}\vspace*{5pt}} 
\setcounter{secnumdepth}{5}

\makeatletter 
\renewcommand\@biblabel[1]{#1}            
\renewcommand\@makefntext[1]%
{\noindent\makebox[0pt][r]{\@thefnmark\,}#1}
\makeatother 
\renewcommand{\figurename}{\small{Fig.}~}
\sectionfont{\sffamily\Large}
\subsectionfont{\normalsize}
\subsubsectionfont{\bf}
\setstretch{1.125} 
\setlength{\skip\footins}{0.8cm}
\setlength{\footnotesep}{0.25cm}
\setlength{\jot}{10pt}
\titlespacing*{\section}{0pt}{4pt}{4pt}
\titlespacing*{\subsection}{0pt}{15pt}{1pt}

\fancyfoot{}
\fancyfoot[RO]{\footnotesize{\sffamily{1--\pageref{LastPage} ~\textbar  \hspace{2pt}\thepage}}}
\fancyfoot[LE]{\footnotesize{\sffamily{\thepage~\textbar\hspace{3.45cm} 1--\pageref{LastPage}}}}
\fancyhead{}
\renewcommand{\headrulewidth}{0pt} 
\renewcommand{\footrulewidth}{0pt}
\setlength{\arrayrulewidth}{1pt}
\setlength{\columnsep}{6.5mm}
\setlength\bibsep{1pt}

\makeatletter 
\newlength{\figrulesep} 
\setlength{\figrulesep}{0.5\textfloatsep} 

\newcommand{\topfigrule}{\vspace*{-1pt}%
\noindent{\color{cream}\rule[-\figrulesep]{\columnwidth}{1.5pt}} }

\newcommand{\botfigrule}{\vspace*{-2pt}%
\noindent{\color{cream}\rule[\figrulesep]{\columnwidth}{1.5pt}} }

\newcommand{\dblfigrule}{\vspace*{-1pt}%
\noindent{\color{cream}\rule[-\figrulesep]{\textwidth}{1.5pt}} }

\makeatother

\vspace{3cm}
\sffamily


\onecolumn
\noindent\LARGE{\textbf{Crystal growth and characterization of the pyrochlore Tb$_2$Ti$_2$O$_7$}} \\
\vspace{0.3cm} \\

\noindent\large{D. Klimm$^{\ast}$\textit{$^{a}$}, C. Guguschev\textit{$^{a}$}, D. J. Kok\textit{$^{a}$}, M. Naumann\textit{$^{a}$}, L. Ackermann\textit{$^{b}$}, D. Rytz\textit{$^{b}$}, M. Peltz\textit{$^{b}$}, K. Dupr\'e\textit{$^{b}$}, M. D. Neumann\textit{$^{c}$}, A. Kwasniewski\textit{$^{a}$}, D. G. Schlom\textit{$^{d,e}$}, and M. Bickermann\textit{$^{a}$}}  \\

\noindent\normalsize{Terbium titanate (Tb$_2$Ti$_2$O$_7$) is a spin-ice material with remarkable magneto-optical properties. It has a high Verdet constant and is a promising substrate crystal for the epitaxy of quantum materials with the pyrochlore structure. Large single crystals with adequate quality of Tb$_2$Ti$_2$O$_7$ or any pyrochlore are not available so far. Here we report the growth of high-quality bulk crystals using the Czochralski method to pull crystals from the melt. Prior work using the automated Czochralski method has suffered from growth instabilities like diameter fluctuation, foot formation and subsequent spiraling shortly after the seeding stage. In this study, the volumes of the crystals were strongly increased to several cubic centimeters by means of manual growth control, leading to crystal diameters up to 40\,mm and crystal lengths up to 10\,mm. Rocking curve measurements revealed full width at half maximum values between 28 and 40$''$ for 222 reflections. The specific heat capacity $c_p$ was measured between room temperature and 1573\,K by dynamic differential scanning calorimetry and shows the typical slow parabolic rise. In contrast, the thermal conductivity $\kappa(T)$ shows a minimum near 700\,K and increases at higher temperature $T$. Optical spectroscopy was performed at room temperature from the ultraviolet to the near infrared region, and additionally in the near infrared region up to 1623\,K. The optical transmission properties and the crystal color are interpreted to be influenced by partial oxidation of Tb$^{3+}$ to Tb$^{4+}$.} \\


\includegraphics[width=0.9\textwidth]{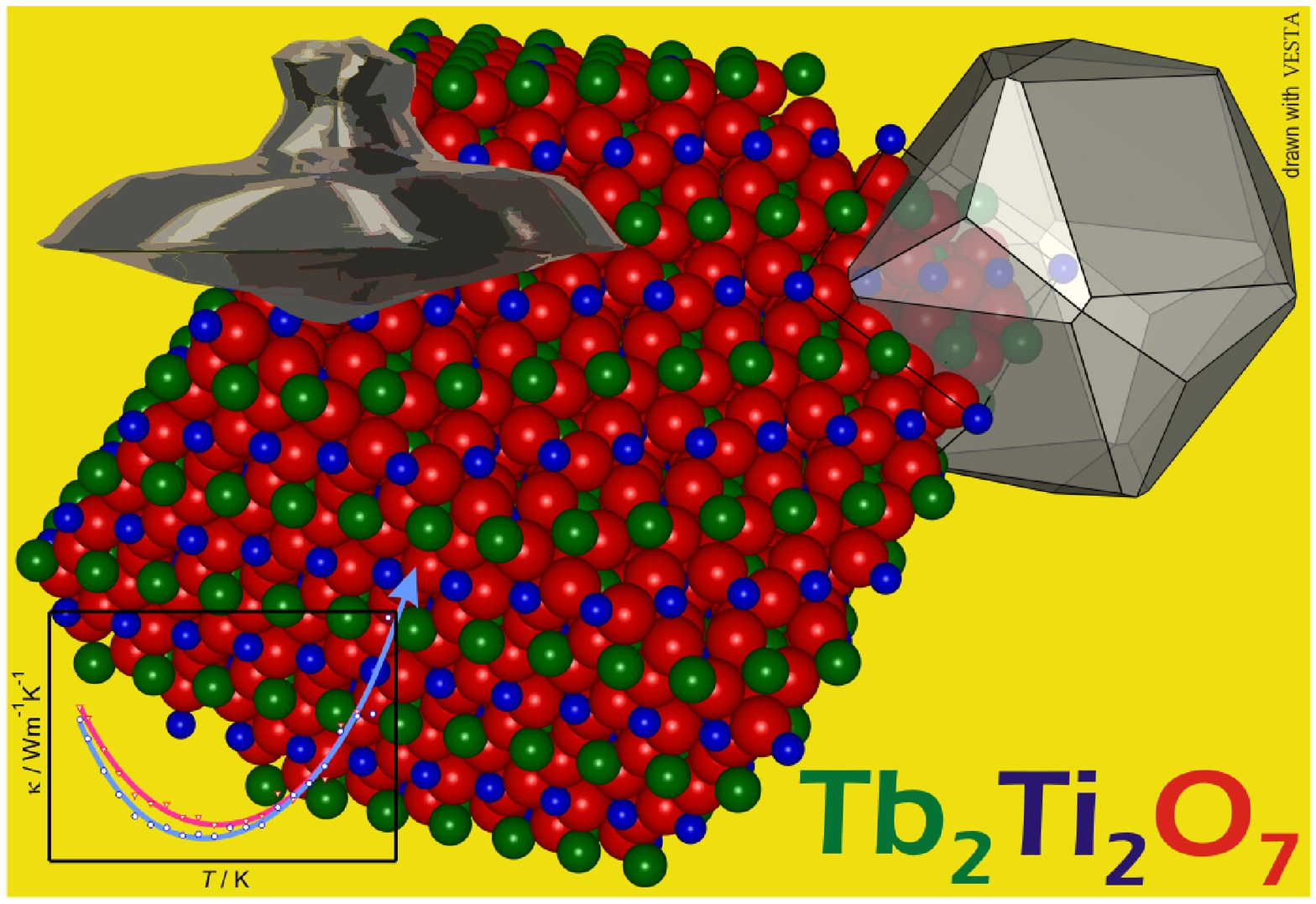}


\renewcommand*\rmdefault{bch}\normalfont\upshape
\rmfamily
\vspace{-1cm}


\footnotetext{\textit{$^{a}$~Leibniz-Institut f\"ur Kristallz\"uchtung, Max-Born-Str. 2, 12489 Berlin, Germany. Fax: 49 6392 3003; Tel: 49 6392 3018; E-mail: detlef.klimm@ikz-berlin.de}}
\footnotetext{\textit{$^{b}$~Forschungsinstitut f\"ur mineralische und metallische Werkstoffe -- Edelstei\-ne/Edel\-metalle -- GmbH (FEE), Struthstr. 2, 55743 Idar-Oberstein, Germany.}}
\footnotetext{\textit{$^{c}$~Leibniz-Institut f\"ur Analytische Wissenschaften, Schwarzschildstr. 8, 12489 Berlin, Germany. Current address: Berliner Glas KGaA, Waldkraiburger Str. 5, 12347 Berlin, Germany}}
\footnotetext{\textit{$^{d}$~Department of Materials Science and Engineering, Cornell University, Ithaca, NY 14853-1501, USA.}}
\footnotetext{\textit{$^{e}$~Kavli Institute at Cornell for Nanoscale Science, Ithaca, NY 14853, USA.}}

\twocolumn

\section{Introduction}

Pyrochlore oxides have the general formula A$_2$B$_2$O$_7$ and crystallize in a cubic structure with space group $Fd\bar{3}m$. They are known for interesting magnetic properties such as giant magnetoresistance\cite{Shimakawa96}. It is generally accepted that the peculiar magnetic properties are related to degeneracy or ``frustration'' for the ground state of the pyrochlore structure where 3D antiferromagnetic order is impossible, resulting in a spin-ice disorder of magnetic moments with nonzero entropy \cite{Bramwell01}. In addition to frustration of magnetic moments, some pyrochlores show structural disorder of the A- and B-sites. For the pyrochlore Tb$_2$Ti$_2$O$_7$, however, it was shown that neither A/B disorder nor oxygen nonstoichiometry play a significant role, making it an almost ideal, disorder-free pyrochlore model substance \cite{Han04}, although minor deviations from the ideal Tb/Ti stoichiometry seem to influence the spin-lattice coupling below 1\,K \cite{Ruminy16}. This compound is also considered to be a prospective Faraday isolator, similar to some other complex cubic terbium oxides or halides \cite{Stevens16}.

Faraday rotators based on Czochralski (Cz) grown Tb$_2$Ti$_2$O$_7$ single crystals were patented recently by some of the current authors \cite{lothar2011terbiumtitanat}. For this prior work, disc-shaped crystals with diameters of 30\,mm and several millimeter in length were grown. Crystals with pronounced $\lbrace111\rbrace$ facets were also reported recently by Guo et al. \cite{Guo16}. They demonstrated crystals with diameter and length of about 20\,mm and 16\,mm, respectively. In the case of crystals grown by the crucible-free floating zone technique, full width at half maximum (FWHM) values of 1080$''$ were reported for the 440 reflection using neutron diffraction \cite{Gardner98}. These crystals had diameters of 3--5\,mm and lengths up to 20\,mm. Until now, no statements on structural quality were published for Cz grown crystals. In the present in work, the growth of bigger single-crystals with superior crystal quality is reported for the first time. High crystalline quality is expected to be relevant especially for optical applications requiring low stress birefrigence.

The pyrochlore structure is host to unusual electronic and magnetic properties --- including spin ice\cite{Bramwell01,Raminez99}, quantum spin ice\cite{Mirebeau05,Bert06,Zhou08,Kimura13,Princep13,Yaouanc13,Tokiwa14}, promising catalysts\cite{Mallat04,Beck06}, and possibly Weyl semimetals\cite{Vafek14}. This makes it of significant interest as a substrate to enable the growth of high quality thin films with the pyrochlore structure, including pyrochlore heterostructures that are predicted to give rise to odd-parity topological superconductivity\cite{She16}. Further, the availability of pyrochlore substrates would enable the strain state in the overlying pyrochlore film to be deliberately tuned to alter its properties. Unfortunately, there are no commercially available pyrochlore single crystal substrates. This has led some to grow epitaxial pyrochlore films on small pyrochlore single crystals grown by the floating-zone method\cite{Bovo14}; substrates with higher structural quality, such as those we report, are a prerequisite to the growth of pyrochlore thin films with higher structural quality.

\section{Experimental}

Stoichiometric melt compositions of Tb$_2$Ti$_2$O$_7$ with a melting temperature\cite{Shcherbakova80} of about $1860^{\,\circ}$C were used for the Cz experiments. As starting materials, dried and mixed powders of Tb$_4$O$_7$ and TiO$_2$ with purities of 99.99\% were used. To optimize the crucible filling process, cylindrical bars were made from the powder mixtures by cold isostatic pressing at 200\,MPa. The crystal growth experiments were performed using a conventional RF-heated Cz setup equipped with a crystal balance. Experiments were performed with manual diameter control under a slightly oxygen enriched argon atmosphere (0.155\,vol\% O$_2$) at ambient pressure. Iridium crucibles (about 60\,mm in diameter and height) embedded in ZrO$_2$ and Al$_2$O$_3$ insulation were used. An actively heated iridium afterheater was placed on top of the crucible. Averaged growth rates were between 2.5 and 30\,mm/h. Tb$_2$Ti$_2$O$_7$ single crystals from preliminary experiments with $\langle111\rangle$ orientation were used as seed material.

Rocking curve X-ray diffraction (XRD) measurements were performed on as-grown $\lbrace111\rbrace$ facets using a high resolution diffractometer (General Electric) with CuK$\alpha_1$ radiation ($\lambda=1.5406$\,\AA). Tb$_2$Ti$_2$O$_7$ 222 Bragg peaks were used for the evaluation of the crystalline quality. For all scans, the collimated beam had a divergence of 11$''$ and the measurement spot covered $\approx10$\,mm length of the sample surface. The spot width depended on the primary beam aperture, which was 0.05\,mm, 0.3\,mm, or 2.0\,mm, respectively.

The specific heat capacity, $c_p$, as a function of temperature $T$ was measured by heat flux differential scanning calorimetry (DSC) with a NETZSCH STA449C. For this purpose, four subsequent heating runs (20\,K/min) from 313\,K to 1573\,K were performed with one piece of a Tb$_2$Ti$_2$O$_7$ crystal in a Pt crucible with a lid in flowing Ar/O$_2$. $c_p(T)$ was obtained by comparison of these DSC curves with analogous curves that were measured with a sapphire standard sample. The first run showed higher experimental scatter and deviated by up to 5\% from the subsequent runs 2--4, probably due to insufficient thermal contact. Therefore, solely the average values of runs 2--4 were used to determine $c_p(T)$.

Measurements of thermal transport were performed with a laser flash apparatus NETZSCH LFA427. For this purpose, sample slices with 1.34\,mm or 1.5\, mm thickness where covered with graphite spray on their top and bottom faces. The sample was heated in flowing N$_2$ to the corresponding temperature step between room temperature and approximately 1380\,K, with a typical difference of 50\,K between steps. At each step the bottom side of the sample was irradiated by three shots of a Nd:YAG laser. Measurement of the top side temperature versus time resulted in a characteristic line shape that was fitted by Mehling's model \cite{Mehling98} for semi-transparent media. This yielded the thermal diffusivity $a$ averaged over the shots. With
\begin{equation}
\kappa = a \, c_p \, \rho   \label{eq:kappa}
\end{equation}
($\rho=6.538$\,g/cm$^3$ -- mass density\cite{Luan11}), the thermal conductivity was obtained by combining $a(T)$ with the experimental $c_p(T)$. $\rho$ was assumed to be constant, which should result in an uncertainty not exceeding a few percent. Note that there should not be any directional dependence to the thermal conductivity of Tb$_2$Ti$_2$O$_7$ because Tb$_2$Ti$_2$O$_7$ is cubic and thermal conductivity is a second-rank tensor\cite{Nye57}.

For the room temperature UV/VIS/NIR spectra, two samples were chemo-mechanically polished to about 1\,mm thickness. One sample was annealed for 12\,h at 1273\,K in air in a muffle furnace. Another sample was cut from a crystal grown at the Forschungsinstitut f\"ur mineralische und metallische Werkstoffe -- Edelsteine/Edelmetalle -- GmbH (FEE) and polished to about 1.5\,mm thickness. The spectra of the samples were recorded using a Perkin-Elmer Lambda~19 spectrometer and are shown in Fig. \ref{TTO_RT_spectra} a). From the measured transmission, $Tr$, the absorption coefficient was calculated using\cite{Schroder06}
\begin{equation}
\alpha=-\frac{1}{d} \ln \frac{\sqrt{(1-R)^{4}+4Tr^{2}R^{2}}-(1-R)^{2}}{2TrR^{2}}
\label{eq:abscoeff}
\end{equation}
where $d$ is the sample thickness and $R$ is the reflectivity.

\begin{figure}
\centering
\includegraphics[height=5.5cm]{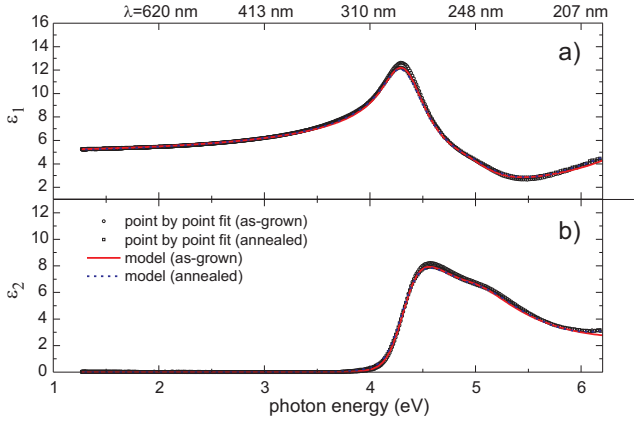}
\caption{Comparison of the model and the point-by-point fitted real (a) and imaginary (b) parts of the dielectric functions of the Tb$_2$Ti$_2$O$_7$ crystals. The onset of strong absorption around the fundamental band gap is around 300\,nm.}
\label{dielectric}
\end{figure}

In order to obtain the reflectivity, a Sentech SE850 ellipsometer was used. The ellipsometric parameters $\Psi$ and $\Delta$ were recorded at room temperature for angles of incidence between 46$^\circ$ and 74$^\circ$ in steps of 7$^\circ$ over the wavelength range from 200 to 970\,nm. By fitting all $\Psi$ and $\Delta$ spectra using a three-layer model, which takes into account the surface roughness via Bruggeman effective medium approximation\cite{Bruggeman35}, the complex dielectric functions of as-grown as well as annealed Tb$_2$Ti$_2$O$_7$ were determined. In the final step, the real and imaginary parts of the dielectric functions obtained by the above procedure were fitted separately for each wavelength without any assumption concerning the line shape, yielding the so-called point-by-point dielectric functions shown in Fig.~\ref{dielectric}. Details of this procedure were reported elsewhere\cite{Neumann16}. The reflectivity was obtained by means of Fresnel coefficients.

High temperature IR spectra were measured for the as-grown and annealed samples in a custom setup where the samples are mounted in a tube furnace (Hesse Instruments HT-19) that can reach up to 2000\,K, giving a maximum possible sample temperature of about 1900\,K. The sample compartment can be flushed with different gases. Light from a high power halogen lamp was passed through the sample in the furnace and into an Acton SpectraPro 2300i spectrograph with an InGaAs diode array. The spectrograph was placed far away from the furnace so most of the diffuse thermal radiation was not recorded\cite{Schwabe03} (the setup is described in detail in Ref. \cite{Kok2015}). Since the setup does not have a reference beam, only intensity in arbitrary units is recorded. This is converted to transmission using
\begin{equation}  
Tr=\frac{I(T)}{I(RT)}\times Tr(\lambda 19)
\end{equation}
where $I(T)$ is the intensity at temperature $T$, $I(RT)$ is the room temperature intensity and $Tr(\lambda 19)$ is the transmission measured with the Lambda~19 spectrometer. From these transmission spectra, the absorption coefficients were calculated using equation \ref{eq:abscoeff}. The reflectivity was extrapolated to match this wavelength range.

\section{Results and Discussion}

\begin{figure}[ht]
\centering
\includegraphics[height=6cm]{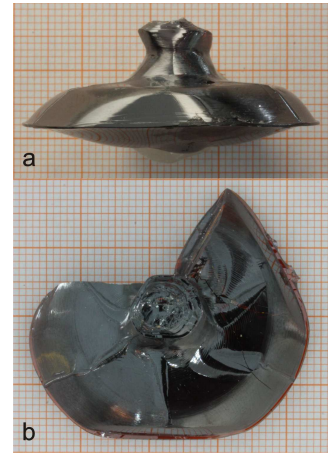}
\caption{As-grown Tb$_2$Ti$_2$O$_7$ Cz single crystal (length 10\,mm, diameter 40\,mm) shown from the side (a) and from the top (b), respectively. The crystal was grown at a rate of about 8\,mm/h. }
\label{fig:x-tal}
\end{figure}

\begin{figure}[ht]
\centering
\includegraphics[height=5cm]{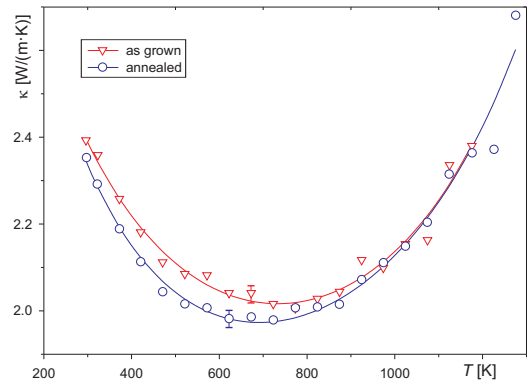}
\caption{Thermal conductivity data $\kappa(T)$ of two Tb$_2$Ti$_2$O$_7$ samples fitted with the function given in equation (\ref{eq:kappa-fit}). Fit parameters are given in table~\ref{tbl:fit-par}.}
\label{fig:kappa}
\end{figure}

Tb$_2$Ti$_2$O$_7$ single crystals with diameters between 30\,mm and 42\,mm and lengths up to 10\,mm (crystal volumes $\geq5.9$\,cm$^3$) were grown by the Cz method. Growth instability resulting from low heat flow through the melt/crystal interface could be suppressed substantially by a combination of manual growth control, moderate to high pulling rates and a sufficient undercooling of the melt. The most critical part of the growth process was found to be the transition to lateral growth during crystal broadening, which sometimes led to the formation of grain boundaries. If at this stage no grain boundaries were formed, it was unlikely that they were formed later on. The as-grown crystals were transparent in bright light and had brownish to dark-red coloration. Unfortunately the low thermal conductivity $\kappa$ (see Fig.~\ref{fig:kappa}) made it necessary to use high temperature gradients to maintain sufficient axial heat flux through the growing crystal. Upon cooling to room temperature this high gradient (directly above the melt surface) usually caused some internal strain. This led to cracking of the crystals into several large parts during handling or sample processing. Cracking could be avoided by cooling the crystals in a lower temperature gradient within the afterheater zone. The crystal from Fig.~\ref{fig:x-tal}a-b cracked only partially during the last stage of growth, due to contact with metal parts of the growth setup, most probably due to local foot generation at the crystal rim.

\begin{figure}[ht]
\centering
\includegraphics[height=4cm]{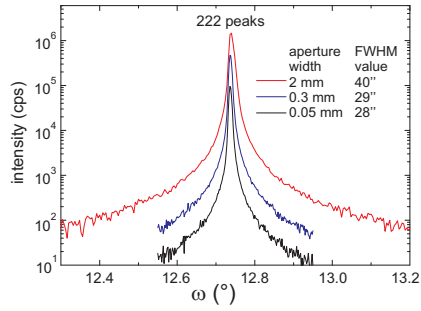}
\caption{X-ray rocking curves of 222 peaks measured on an as-grown $\lbrace111\rbrace$ facet using three different primary beam apertures. The crystal was grown at a rate of about 2.5\,mm/h.}
\label{fig:rocking}
\end{figure}

Initially, heterogeneously nucleated boules were grown on Ir tubes, which resulted in multicrystalline growth during the increase of the crystal diameter. Nonetheless, these boules also contained large single crystalline parts suitable for high-quality seed and sample preparation. Single crystals with high structural quality were obtained using $\langle111\rangle$ oriented seeds and moderate growth rates between 2.5 and about 8\,mm/h. For the single crystalline boules, structural quality was investigated for a crystal grown at a rate of 2.5\,mm/h. The results of the X-ray rocking curve measurements, performed on an as-grown facet (not polished), show that the FWHM values lie between 28 and 40$''$ depending on the aperture width of the diffractometer's X-ray source (Fig.~\ref{fig:rocking}). Most probably, these slight differences of FWHM values can be explained by a minor surface curvature of the sample or by residual stress, since the crystal was cooled down directly above the melt. The results of the rocking curve measurements show that a remarkably high crystal quality can be achieved using the Cz method. This is in contrast to the reported structural quality of Tb$_2$Ti$_2$O$_7$ crystals grown under very high thermal gradients using the optical floating zone method \cite{Gardner98}.

The structural quality is also higher than for Cz grown SrTiO$_3$ crystals showing similar growth behavior \cite{Guguschev15d}. Potentially, lower thermal gradients resulting from the $\approx200$\,K lower melting temperature here led to higher crystal quality. Additionally, a slightly higher transmissivity in the near infrared region at the moderate growth temperatures and presumably a slightly increased thermal conductivity at high temperatures led to a higher total heat transport through the crystal and to a more stable growth of Tb$_2$Ti$_2$O$_7$ than SrTiO$_3$. Since heat transport is still very low (compared to other oxide crystals like corundum ($\alpha$-Al$_2$O$_3$), the achievable crystal length with stable growth behavior is limited to 1--2\,cm.

\begin{figure*}[ht]
\centering
\includegraphics[height=5cm]{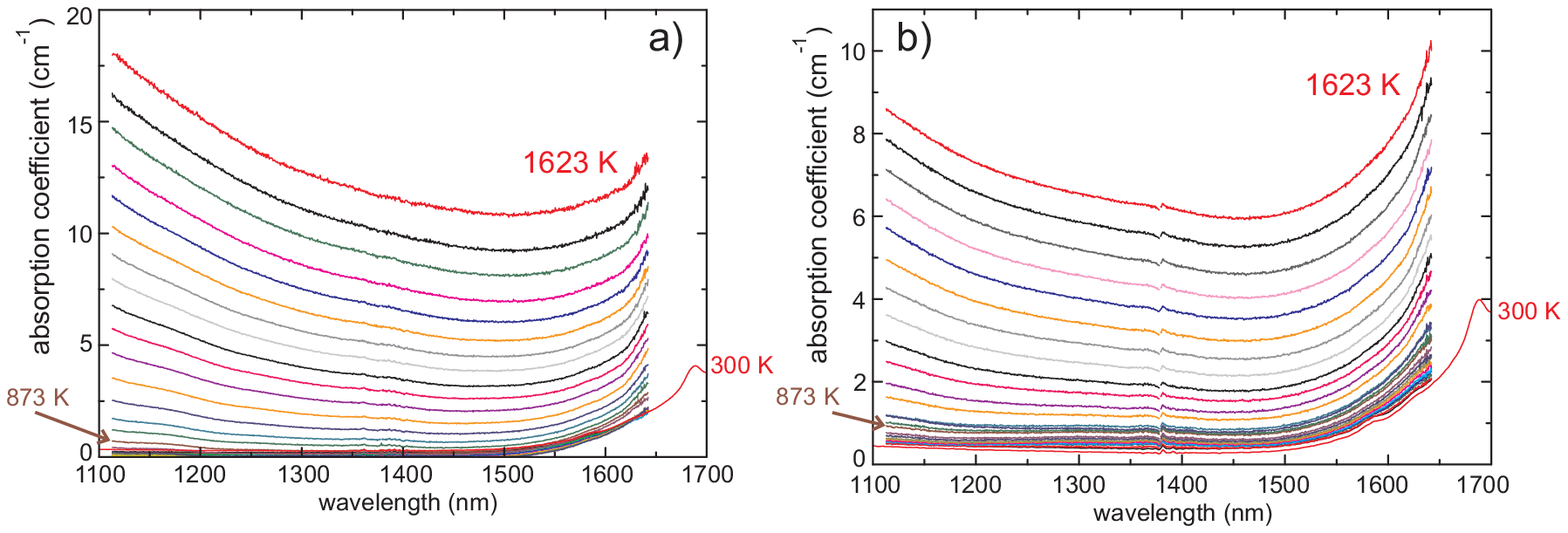}
\caption{IR high temperature spectra of a) TTO~6-1 annealed and b) TTO~6-1 as-grown, measured during cooling from 1623\,K to RT. The feature near 1380\,nm is an artefact.}
\label{IR_HT_Spectra}
\end{figure*}

Also in contrast to SrTiO$_3$ \cite{Guguschev15c}, the infrared absorption of Tb$_2$Ti$_2$O$_7$ after annealing in strongly oxidizing atmosphere (1273\,K in air, $p_{\text{O}_2}=212$\,mbar) is higher than after growth in low oxygen partial pressure (1.6\,mbar, see Fig.~\ref{IR_HT_Spectra}), i.e. growth in a highly oxidizing atmosphere is not helpful to stabilize the growth process. Perhaps there is some potential to increase the lengths of the crystals by application of a very high axial thermal gradient, but this will ultimately lead to a lower crystalline quality and cracking of the crystals.

\begin{figure*}
\centering
\includegraphics[height=5cm]{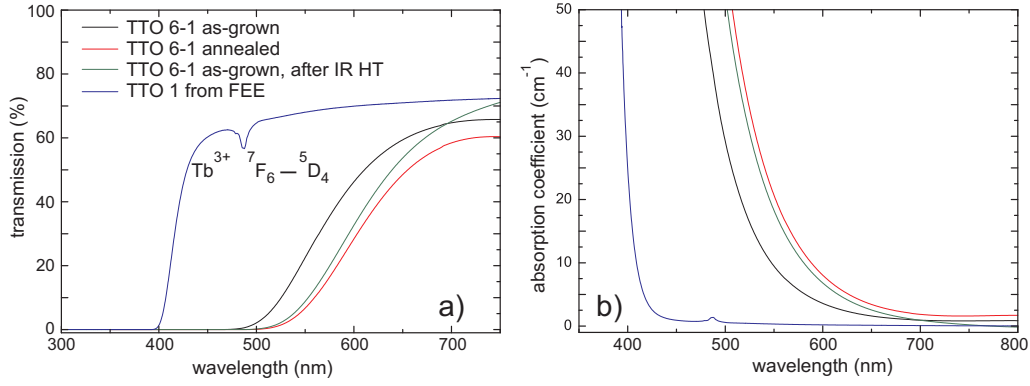}
\caption{a) Transmission and b) absorption coefficient spectra of the Tb$_2$Ti$_2$O$_7$ crystals. The peak shown by the FEE crystal at about 490\,nm is probably the $^7$F$_6$--$^5$D$_4$ transition of the Tb$^{3+}$ ions \cite{Yoshida11,Villora11}. The onset of absorption shifts to longer wavelength after annealing.}
\label{TTO_RT_spectra}
\end{figure*}

Room temperature transmission spectra of the different Tb$_2$Ti$_2$O$_7$ samples are shown in Fig.~\ref{TTO_RT_spectra}. Remarkably, a sharp absorption edge is observed only for sample TTO~1 grown by some of the current authors (L.A, D.R., M.P., \& K.D.) in another Cz setup, with lower oxygen partial pressure in the atmosphere (pure nitrogen without oxygen admixture) and from purer material (99.999\%\ instead of 99.99\%) than crystal TTO~6-1. The peak shown by this crystal near 490\,nm is the $^7$F$_6$--$^5$D$_4$ transition of the Tb$^{3+}$ ions \cite{Yoshida11,Villora11}. The broad absorption convoluted with the band gap is shifted further to longer wavelengths for the annealed sample.

Thus one can assume that the partial oxidation of Tb$^{3+}$ to Tb$^{4+}$ is responsible for the coloration of the crystals rather than the presence of Ti$^{3+}$. The stability limits of the different terbium oxides as a function of oxygen fugacity are calculated with FactSage \cite{FactSage7_0} to be
\begin{equation}
\text{TbO}_2 \xlongleftrightarrow{412-582~ \text{K}} \text{Tb}_6\text{O}_{11} \xlongleftrightarrow{498-685~ \text{K}} \text{Tb}_7\text{O}_{12} \xlongleftrightarrow{679-963 ~\text{K}} \text{Tb}_2\text{O}_3 , \label{eq:terbium}
\end{equation}
where the lower temperature represents the stability limit for a virtually oxygen free atmosphere ($\log[p_{\text{O}_2}]=-5$), and the higher number is valid in air ($\log[p_{\text{O}_2}]=-0.678$). From eqn. (\ref{eq:terbium}) it is obvious that at least the formation of Tb$_7$O$_{12}=$2\,Tb$_2$O$_3\cdot$3\,TbO$_2$ is feasible under all accessible experimental conditions. The stability limit in a virtually oxygen free atmosphere, however, is about 200\,K lower, compared to air. This raises the chance that Tb$_2$O$_3$, and hence pure Tb$^{3+}$, is frozen in and held metastable also down to room temperature. Tb$^{4+}$ is smaller than Tb$^{3+}$ (90 vs. 106.3\,pm)\cite{Shannon76}, but still larger than Ti$^{4+}$ (74.5\,pm) and hence Tb/Ti site interchange seems unrealistic, in agreement with recent experimental results\cite{Han04}. The equilibrium (\ref{eq:terbium}) can also explain the higher absorption of air annealed Tb$_2$Ti$_2$O$_7$ (Fig.~\ref{IR_HT_Spectra}), and is in agreement with recent experiments on terbium scandate TbScO$_3$, were a mass gain resulting from partial Tb$^{4+}$ formation was observed upon air annealing \cite{Uecker08}. Also Tb$_2$Ti$_2$O$_7$ is very dark if grown under high $p_{\text{O}_2}=0.4$\,MPa \cite{Li13b}. It can be calculated that under the given experimental conditions titanium occurs only as Ti$^{4+}$ \cite{FactSage7_0}.

During the high temperature IR measurements an annealing effect was observed for the as grown crystal (see Fig. \ref{TTO_RT_spectra}), probably due to air leaking into the furnace. Because of that, only the cooling spectra are evaluated. The high temperature spectra of both samples are qualitatively very similar, but the absorption coefficients are much higher for the annealed sample (see Fig. \ref{IR_HT_Spectra}). The rise of the absorption coefficient with temperature is fairly monotonic (see Fig. \ref{IR_HT_Spectra}) and the shape of the spectra does not match free carrier absorption, which is proportional to $\lambda^2$. There is an increase in absorption at the long wavelength edge of the spectra, but this is just the edge of the peak around 1700\,nm which is shown in the room temperature spectra. It appears that less than about 10$^{17}$\,cm$^{-3}$ free electrons are present, since this is usually the required approximate concentration for a significant absorption \cite{Schroder06}. With rising temperature, the absorption on the low wavelength side increases faster than in other parts of the spectrum (see Fig. \ref{IR_HT_Spectra}). This might be caused by the shifting or broadening of the Tb$^{4+}$ absorption that obscures the band gap.

\begin{figure}[ht]
\centering
  \includegraphics[height=5cm]{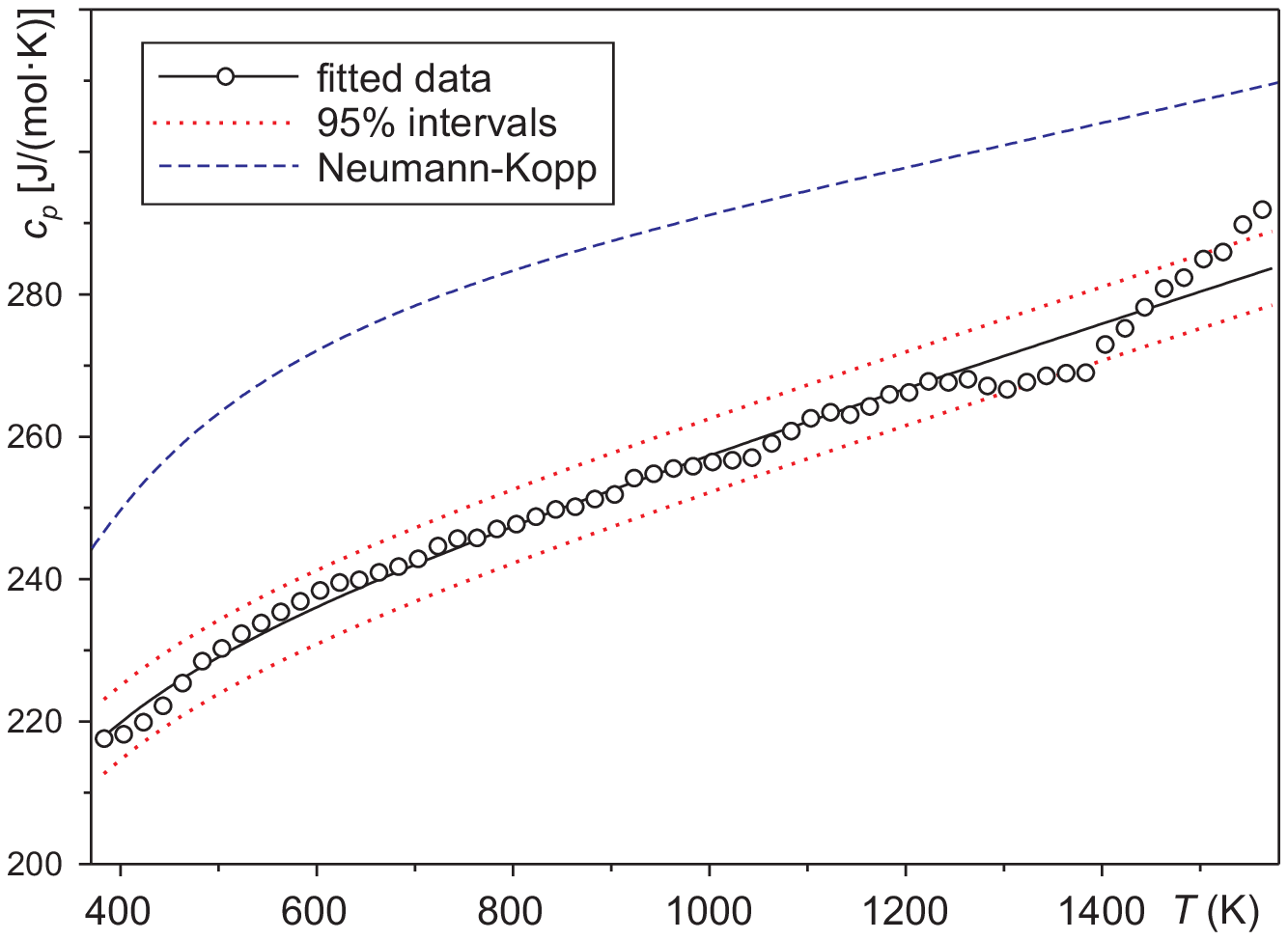}
  \caption{Measured $c_p$ data, together with a fit to the function in equation (\ref{eq:cp}) and 95\% prediction bands. Also shown for comparison are $c_p(T)$ data calculated according to the Neumann-Kopp rule described by equation (\ref{eq:Neumann}) from Tb$_2$O$_3$ and TiO$_2$ data.}
  \label{fig:c_p}
\end{figure}

The DSC measurements did not show significant hints of any phase transformations over the experimental range $383$\,K $\leq T\leq1573$\,K. Larger deviations from smooth behavior of $c_p(T)$ at $T>1300$\,K (Fig.~\ref{fig:c_p}) are not reproducible and result from experimental scatter and some drift because at high $T$ the DSC signal is not solely influenced by thermal conduction. The experimental data could be fitted satisfactory by a simple expression
\begin{equation}
c_p = a + bT + c/T^2   \label{eq:cp}
\end{equation}
with parameters $a=215.87, b=0.0436, c=-2.154574\times10^6$, $c_p$ given in J/(mol\,K). Often the Neumann-Kopp rule gives a good approximation for the $c_p(T)$ functions of complex oxides as the sum of the $c_p(T)$ functions of their constituents. In Fig.~\ref{fig:c_p} the $c_p(T)$ data are fitted to equation (\ref{eq:cp}) and are compared to the sum of the heat capacities of the constituents following the formation reaction
\begin{equation}
\text{Tb}_2\text{O}_3 + \text{2\,TiO}_2 \longrightarrow \text{Tb}_2\text{Ti}_2\text{O}_7  \label{eq:Neumann}
\end{equation}
that is shown as a dashed line. The disagreement can be attributed to the formation energy of reaction (\ref{eq:Neumann}), which is neglected by the Neumann-Kopp rule. 

Because Tb$_2$Ti$_2$O$_7$ is cubic, no anisotropy may occur for the second-rank $\kappa$ tensor\cite{Nye57}. Hence, $\kappa(T)$ is represented by a single value for every $T$. The function $\kappa(T)$ was measured for several sample slices, and the results were similar for all of them. Figure~\ref{fig:kappa} shows experimental points for two samples. Typically near $T>1200$\,K the measurements had to be stopped because the experimental scatter became too large and soon the detector signal of the laser flash apparatus diverged. This instability resulted obviously from damage of the graphite absorption layers, probably by chemical reaction of the Ti$^{4+}$ ions with the strongly reducing graphite.

For many materials the thermal conductivity drops at high $T\gg T_\Theta$ ($T_\Theta$ -- Debye temperature) following a $\kappa\propto T^{-1}$ law as a result of phonon scattering (umklapp process) \cite{Slack62}, especially for isolators such as most oxides. Hofmeister \cite{Hofmeister06} found for the laser flash thermal diffusivity of several garnets expressions of the type
\begin{equation}
a=A+BT+CT^2        \label{eq:Hofmeister}
\end{equation}
($A,B,C$ -- constants) to be well suited for fitting the thermal dependence. It should be noted that the low variation of $c_p$ at high $T$ allows similar descriptions for $a(T)$ and $\kappa(T)$. The experimental data in Fig.~\ref{fig:kappa}, however, showed a clear rise of $\kappa$ for large $T>700$\,K, which both models fail to explain.

More recently Glassbrenner \cite{Glassbrenner64} performed thermal conductivity measurements for germanium and silicon up to their melting points $T_\text{f}$. They showed that for all temperatures the major contribution to $\kappa$ is produced by phonons; but at sufficiently high $T\approx700$ or 1000\,K, respectively, a significant share of heat is transported by electrons. For the lattice thermal resistance $W=\kappa^{-1}$ a dependence of the type described by equation (\ref{eq:Hofmeister}) was found, where the quadratic term is attributed to 4-phonon processes. For temperatures close to $T_\text{f}$, an almost flat $\kappa(T)$ dependence was measured for Si, and for Ge even a growth of $\kappa(T)$ was observed starting ca. 100\,K below $T_\text{f}$. Heat transport by free carriers does not, however, seem realistic in the present case, because the optical measurements showed that the electron density is certainly low, under 10$^{17}$\,cm$^{-3}$. On the other hand, the opacity of dielectric materials typically increases with $T$, diminishing thermal transport by radiation\cite{Aronson70}. This is confirmed by the current laser flash measurements where the detector signal shows that the instantaneous heat transport (by radiation) drops with $T$. It should be noted that also in the inverse spinel MgGa$_2$O$_4$, which is another oxide crystal with low carrier density, $\kappa$ was found to grow at high $T\approx1200$\,K\cite{Schwarz15}.

The limited range of experimental points and the higher experimental scatter above the minima in the $\kappa(T)$ functions in Fig.~\ref{fig:kappa} does not allow an accurate determination of the temperature dependence of the process that results in a rising $\kappa(T)$. Here, a linear behavior was found to describe the experimental data well; hence in the empirical formula
\begin{equation}
\kappa = \frac{1}{A + B T + C T^2} + D T   \label{eq:kappa-fit}
\end{equation}
the first term is assumed to describe the thermal transport by phonons ($\propto W^{-1})$, and the second (linear) term describes well an additional high-$T$ term that might be related to heat transport by charge carriers (electrons or holes, respectively).

\begin{table}[ht] 
\small
\caption{\ Fit function parameters from equation (\ref{eq:kappa-fit}) for the $\kappa(T)$ functions in Fig.~\ref{fig:kappa}.}
\label{tbl:fit-par}
\begin{tabular*}{0.40\textwidth}{@{\extracolsep{\fill}}lll}
\hline
Parameter & as grown               & annealed \\
\hline
$A$ (W$^{-1}$\,m\,K)             & $0.2704$               & $0.2334$ \rule{0mm}{3mm}   \\
$B$ (W$^{-1}$\,m)                & $6.3300\times10^{-4}$  & $8.8523\times10^{-4}$ \\
$C$ (W$^{-1}$\,m\,K$^{-1}$)      & $-4.1922\times10^{-7}$ & $-5.4826\times10^{-7}$ \\
$D$ (W\,m$^{-1}$\,K$^{-2}$)      & $7.0879\times10^{-5}$  & $3.7419\times10^{-4}$ \\
    \hline
  \end{tabular*}
\end{table}

\section{Conclusions}

High quality single crystals of Tb$_2$Ti$_2$O$_7$ were grown using the Czochralski method. The growth conditions used allowed the growth of disc-shaped single crystals up to 10\,mm in length and up to 40\,mm in diameter. Unfortunately, slow diameter enlargement and stable long-term growth was not possible due to the poor heat transport through the growing crystal. Cracking of the crystals upon cooling was largely prevented by cooling the crystals with a lower temperature gradient. 

The temperature dependence of the thermal conductivity is atypical above $700-900$\,K, because it shows a minimum value around 2\,W/(m$\cdot$K) and is slightly larger for lower and higher $T$. The range of minimum thermal conductivity corresponds to the temperature range were optical transmission starts to drop. Besides these intrinsic properties, optical transmission is altered remarkably if the Tb$^{3+}$ ions in the material are oxidized partially to Tb$^{4+}$.

Despite the fact that high quality crystals were grown by the Czochralski method, it is rather unlikely that Tb$_2$Ti$_2$O$_7$ single crystals can be produced on an automated industrial scale. It seems to be more appropriate to use the edge-defined film-fed growth (EFG) method, which is proven to permit excellent control of the growth processes for a variety of materials e.g. for rutile\cite{Machida94}, cerium aluminate\cite{Arhipov15}, rare-earth orthovanadates\cite{Epelbaum99}, and strontium titanate\cite{Guguschev15d}. For all of these oxides, Czochralski growth was less successful since pronounced growth instabilities occurred. These instability issues were often triggered by poor heat transport (low infrared transmissivity and/or low thermal conductivity) through the growing crystals.





\section*{Acknowledgements}

The authors are grateful to M. Br\"utzam, E. Thiede, M. Imming-Friedland, and M. Rabe for technical assistance and material preparation. We thank D. Siche for reading the manuscript.


\providecommand*{\mcitethebibliography}{\thebibliography}
\csname @ifundefined\endcsname{endmcitethebibliography}
{\let\endmcitethebibliography\endthebibliography}{}

\end{document}